\input{aipcheck}

\documentclass[
    ,final            
  ]
  {aipproc}

\layoutstyle{8x11double}

\begin{document}

\title{Anomalous Soft Photons associated with Hadron Production in String Fragmentation}

\classification{11.10.Kk 11.15.Tk 13.66.-a 24.85.+p}
\keywords {String fragmentation, soft photon production, hadron production }

\author{Cheuk-Yin Wong }
{
  address={Physics Division, Oak Ridge National Laboratory, 
Oak Ridge, TN 37831}
}

\begin{abstract}

\vspace*{0.2cm}
The bosonized QCD2+QED2 system for quarks with two flavors contains
QCD2 and QED2 bound states, with an isoscalar photon at about 25 MeV
and an isovector $(I$=1,$I_3$=0) photon at about 44 MeV.
Consequently, when a quark and an antiquark at the two ends of a
string pulls apart from each other at high energies, hadrons and soft
photons will be produced simultaneously in the fragmentation of the
string.  The production of the QED2 soft photons in association with
hadrons may explain the anomalous soft photon data in hadron-hadron
collisions and $e^+$-$e^-$ annihilations at high energies.
\vspace*{0.1cm}

\end{abstract}

\maketitle


\section{Introduction}
Anomalous soft photons are low-$p_T$ photons ($p_T$<60 MeV) produced
in excess of what is expected from electromagnetic bremsstrahlung.
They are found to possess the following interesting properties:

\begin{enumerate}

\item Anomalous soft photons are produced 
in $K^+p$ \cite{Chl84}-\cite{Bel97}, $\pi^-p$ \cite{Ban93,Bel02},
$\pi^+p$ \cite{Bot91,Bel97}, and $pp$ \cite{Bel02a} collisions, and in
$e^+e^-$ annihilations \cite{DEL06}-\cite{Per09} at high
energies.  They occur only in association with hadron production.
They are absent in $e^+$+ $e^-$$\to$$ Z^0$$ \to$ $ \mu^+$+ $\mu^-$
in which there is no hadron production \cite{DEL06,DEL08}.

\item The transverse momenta of the anomalous soft photons are found
  to be in groups, at approximately 5, 15, and 50
  MeV \cite{Bel02a,DEL10,Won10}.

\item
The soft photon yield is proportional to the hadron particle
yield \cite{DEL10}.

\item The soft photons are associated more with neutral hadron
  production than with charged hadron production \cite{DEL10}.

\end{enumerate}

Previously, many different theoretical models have been put forth to
explain the phenomenon \cite{Van89}.  However, a complete
understanding is still lacking. 

We note that in the flux tube environment, both color charge
oscillations and electric charge oscillations are quantized, and the
system contains QCD2 and QED2 bound states.  We would like to propose
that these bound hadron and soft photon states will be produced
simultaneously in the process of string fragmentation, when a quark
pulls away from an antiquark at high energies.  The production of
these soft photons in association with the production of hadrons may
explain the anomalous soft photon phenomenon \cite{Won10}.

\section{Approximate Compactification of QCD4+QED4 into QCD2+QED2 }

To investigate the simultaneous production of QCD and QED quanta, we
 consider the U(3) group that is the union of the color SU(3) and the
 electromagnetic U(1) subgroups with coupling constants $g_f^\alpha$
 that depend on the U(3) generator index $\alpha$ and the quark favor
 $f$ in QCD4+QED4,
\begin{eqnarray}
\label{qcdcc} 
g_u^{\{1,..,8\}}=g_d^{\{1,..,8\}}=g_{\rm QCD4}, {\rm~~for~~QCD},
\end{eqnarray}
\vspace*{-0.4cm}
\begin{eqnarray} 
\label{qedcc}
g_u^0=-e_u=-Q_u e, ~~~g_d^0=-e_d=-Q_d e,   {\rm~for~~QED},
\end{eqnarray}
where $Q_u=2/3$, and $Q_d=-1/3$.  In high energy particle production
processes, the dominance of the longitudinal motion of the leading
quark and antiquark and the presence of transverse confinement lead to
the formation of a flux tube that can be approximated as a string.  As
shown in \cite{Won10}, the dynamics of the system of quarks and gauge
fields in QCD4+QED4 can then be approximately compactified into the
dynamics of QCD2+QED2, in which quarks acquire a transverse mass $m_T$
that encodes the information of the flux tube radius $R_T$
\cite{Won10}.  The coupling constants in 2D and 4D space-time are then
related approximately by \cite{Won10}
\begin{eqnarray}
\label{est}
g_{\rm 2D}^2 \sim  \frac{g_{\rm 4D}^2}{\pi R_T^2}.
\end{eqnarray}
For QCD and QED states, the isospin symmetry remains a good symmetry
for QCD but not for QED.  The four $(I$,$I_3$) QED states split apart.
We study only neutral $I_3$=0 bound states: the isoscalar
$(I$=0,$I_3$=0) state and the isovector $(I$=1,$I_3$=0) state, in the
strong coupling limit in which $|g_{\rm QCD2}|$$>>$$ m_T$.

\section{Bound states in QCD2+QED2}

The best method to search for bound states is by bosonization
\cite{Col75}.  We use non-Abelian bosonization \cite{Wit84} for U(3)
interactions, and Abelian bosonization for flavor degrees of freedom.
The bosonization program consists of introducing boson fields to
describe elements of the U(3) group and showing subsequently that
these boson fields lead to stable boson states with finite or zero
masses.  Not all degrees of freedom available to the bosonization
technique in U(3) will lead to good boson states with these desirable
properties. We therefore search for stable bosons by varying only the
amplitude of the boson field in color space, keeping the orientation
in color space fixed \cite{Won10}.  This means that we limit the U(3)
generator index $\alpha$ in Eqs.\ (\ref{qcdcc}) and (\ref{qedcc}) to
be $\alpha=$ 0 and 1.

From bosonization we find that QCD2+QED2 contains bound boson states
because gauge field oscillations lead to quark color and electric
charge density oscillations, and through the Maxwell equations the
quark color and electric charge density oscillations in turn lead to
gauge field oscillations.  The self-consistency of gauge field
oscillations and the induced quark charge density oscillations lead to
an equation of motion for the gauge field oscillations in the form of
a Klein-Gordon equation characterized by isospin-dependent masses
\cite{Col75,Sch62}.

As shown in \cite{Won10},  the mass square $(M_I^\alpha)^2$
of the stable boson in QCD2+QED2 for a state with quantum numbers 
$(I$,$I_3$=0) is given by 
\begin{eqnarray}
\label{pot}
(M_I^\alpha)^2
=\left ( \frac{g_{u(2D)}^\alpha+(-1)^I g_{d(2D)}^\alpha}{\sqrt{2\pi}} \right )^2
+ \frac{2}{3-\alpha} e^\gamma m_T \mu,
\end{eqnarray}
where the index $\alpha$=0 corresponds to QED2, $\alpha$=1 corresponds
 to QCD2, $\gamma$=0.5772 is the Euler constant, and $\mu$ is the mass
 scale for the interaction in question.

\section{Hadron and Photon Masses in QCD2+QED2 with two Flavors}

For QCD2 and QED2 in the flux tube, the coupling constants are related
to the 4D constants as given in Eq.\ (\ref{est}) by
\begin{eqnarray}
\label{R1}
g_{\rm QCD2}^2 \sim 
\frac{4\alpha_s}{R_T^2},
{\rm~~and~~}
e_{\rm QED2}^2 \sim 
\frac{4\alpha}{R_T^2},
\end{eqnarray}
where $\alpha_s$=$g_{\rm QCD4}^2/4\pi$ and $\alpha$=$e_{\rm
 QED4}^2/4\pi$=1/137.  The flux tube radius $R_T$ can be determined
 from the root-mean-squared transverse momentum of produced hadrons as
\begin{eqnarray}
\label{R2}
R_T \sim \frac{1}{\sqrt{\langle p_T^2\rangle_\pi } }.
\end{eqnarray}
The measurement of the $\pi^0$ spectra in $Z^0$ hadronic decay gives
$\sqrt{\langle p_T^2\rangle_\pi }$=0.56 GeV in the reaction plane
\cite{Bar97} and thus the flux tube has a radius $R_T$$\sim$0.35 fm.
For the strong coupling constant at this energy, we shall take
$\alpha_s=0.316$, which leads from Eq.\ (\ref{R1}) to the string
tension coefficient  $b$=${g_{\rm QCD2}^2}/{2}$=0.2 GeV$^2$ \cite{Won09,Won09a}.

With these coupling constants and Eq.\ (\ref{pot}), the QCD2 and QED2
boson masses can be determined as shown in Table I.  In the massless
quark limit, the QCD2 isovector hadron state is massless and lies
lower than the isoscalar hadron state at 505 MeV, whereas the ordering
is opposite for the QED2 states, with an isoscalar photon at 12.8 MeV
and an isovector $I_3$=0 photon at 38.4 MeV.
  
\begin{table}[h]
\caption { Meson and photon masses with    $I_3$=0 in a tube  }
\begin{tabular}{|c|c|c|c|}
\cline{3-4} \multicolumn{2}{c|}{} & QCD2 & QED2 
       \\ \hline 
\multicolumn{2} {|c|} {Coupling Constant (MeV)}
& $g_{\rm QCD2}$=632.5  & $e_{\rm QED2}$=96        \\ \cline{1-4} 
massless quarks  &  I=0   
&  504.6 MeV &  12.8 MeV 
 \\  
        & I=1 
&  0         &  38.4 MeV 
 \\ \cline{1-4}
$\mu$=$m_T$=400 MeV  &  I=0 
&  734.6 MeV &  
        \\  
             & I=1   
&  533.8 MeV &  
           \\ \cline{1-4}  
 $m_T$= 400 MeV  &  I=0 
&            &  25.3 MeV 
 \\  
$\mu$=$m_q$=1 MeV &  I=1   
&                 &  44.1 MeV 
  \\ \cline{1-4}  
\end{tabular}
\end{table}

The value of $m_T$ can be estimated from $\sqrt{\langle p_T^2\rangle_\pi }$.
Because a pion is a quark-antiquark composite, we obtain the quark
transverse mass $m_T$=$\sqrt{\langle p_T^2\rangle_\pi /2}$=0.4 GeV.

The mass scale $\mu$ depends on the bosonization of the scalar
density, which diverges in perturbation theory and has to be
renormalized and renormal-ordered again in an interacting theory
\cite{Col75}.  The mass scale $\mu$ therefore depends on the interaction.
For QCD2, the strong confining interaction dominates and leads to
transverse confinement with a quark transverse mass $m_T$.  It is
reasonable to take the mass scale $\mu$ in QCD2 to be the same $m_T$
characterizing the flux tube transverse confinement.  This $\mu$ value
gives a QCD2 isovector hadron mass of 0.534 GeV and isoscalar hadron
mass of 0.735 GeV (Table I) that agree approximately with the observed
average isovector hadron transverse mass $m_{\pi T}$=0.577 GeV and
isoscalar hadron transverse mass $m_{\eta T}$$\sim$0.824 MeV.

For QED2, we envisage that the scalar density ${\bar \psi} \psi$ that
diverges in perturbation theory has to be renormalized in a free
theory in which the quarks are characterized by the quark current
masses.  The appropriate mass scale $\mu$ for QED2 is therefore the
quark current masses, of order 1 MeV.  The values of the QED2 boson
masses obtained with $m_T$=0.4 GeV and $\mu$=1 MeV are given in Table
I, which gives an isoscalar photon of about 25 MeV and an isovector
photon of about 44 MeV, in approximate agreement with observed soft
photon $p_T$ spectra at approximately 15 and 50 MeV
in \cite{Bel02a,DEL10,Won10}, except for the groups of soft photons
with $p_T$$\sim$5 MeV whose origin remains unknown.

\section{Simultaneous production of QED2 Photons and QCD2 Hadrons}

In hadron-hadron collisions and $e^+$-$e^-$ annihilations at high
energies, a string is formed between a quark and an antiquark (or
diquark).  When the quark and the antiquark pull apart from each other
in the string fragmentation process, the QCD vacuum is so polarized
that hadrons are produced by the strong QCD field, with the rapidity
distribution of the produced hadrons in the form of a plateau, as
discussed by Casher, Kogut and Susskind \cite{Cas74}, and analyzed
phenomenologically in \cite{Won09,Won09a}.

The receding quark and antiquark as well as quarks in the vacuum also
carry electric charges and they interact with electromagnetic
interactions.  Simultaneous with the QCD interaction and the
production of QCD hadrons, the quark and the antiquark generate a
strong electric field between them in the flux tube that can produce
QED photons, if they are stable bosons of the QCD2+QED2 system.

QCD2 hadrons and massive QED2 photons have been found to be stable
quanta in the QCD2+QED2 system.  The isoscalar photon mass
$M_{I=0}^\gamma$ is about 25 MeV and the isovector $I_3$=0 photon mass
$M_{I=1}^\gamma$ is about 44 MeV for a flux tube of radius of 0.35 fm.
The same receding quark and antiquark source will produce both QCD2
hadrons and QED2 photons simultaneously.  The production process
corresponds to the excitation of the QCD2+QED2 vacuum due to the color
and electric charge oscillations generated during the string
fragmentation process.

The simultaneous production of hadrons and photons explains why
anomalous photons are present only in association with hadron
production.  As the source and the production mechanism of both
hadrons and photons are the same, the number of produced hadrons and
QED2 photons are therefore proportional to each other, on an
event-by-event basis, as observed by the DELPHI
Collaboration \cite{DEL10}.

We envisage that after a massive QED2 photon is produced it emerges
from the interacting region to the outside non-interacting region,
with an adiabatic expansion of the flux tube.  The massive QED2 photon
will evolve into a massless QED4 photon adiabatically, with the photon
mass in QED2 turning into the transverse momentum in QED4.  We
therefore expect that a produced isoscalar photon will lead to a
photon with an average transverse momentum of about 25 MeV, and an
isovector $I_3$=0 photon with an average transverse momentum of about
44 MeV.  These transverse momenta fall within the domain of soft
photon transverse momenta observed in $Z^0$ hadronic decay
\cite{DEL10} and hadron-hadron collisions \cite{Bel02a}.

The present model predicts that the isoscalar photon mass is lower
than the isovector photon mass. Consequently, the production of
isoscalar photons is more likely than isovector photons. In contrast,
the QCD isoscalar meson mass is greater than the isovector meson mass,
the production of isoscalar mesons is less likely than that of
isovector mesons.  As a consequence, the ratio $N^\gamma$/$N_{\rm
neutral}$ is much greater than the ratio $N^\gamma$/$N_{\rm charged}$,
as observed by the DELPHI Collaboration \cite{DEL10}.

In conclusion, in the flux tube string-like environment formed after a
high-energy hadron-hadron collision or $e^+$-$e^-$ annihilation, the
QCD2+QED2 system contains stable QCD2 hadrons and QED2 photons.  These
bosons will be produced simultaneously when a quark pulls away from an
antiquark (or diquark) at high energies during the fragmentation of
the string. They may lead to the anomalous soft photons observed in
hadron-hadron collisions and e$^+e^-$ annihilations at high energies.


\vspace{-0.40cm}
\begin{theacknowledgments}
\vspace{-0.20cm}
The research was sponsored by the Office of Nuclear Physics,
U.S. Department of Energy.

\end{theacknowledgments}

\bibliographystyle{aipproc}   
\vspace{-0.40cm}



\begin{thebibliography}{200}

\bibitem{Chl84}
P. V. Chliapnikov $et~al.$, Phys. Lett. {\bf B141}, 276
(1984).

\bibitem{Bot91}
 F. Botterweck $et~al.$, Z. Phys. {\bf C51}, 541 (1991).

\bibitem{Bel97}
A. Belogianni $et~al.$,Phys. Lett. {\bf B408}, 487 (1997).


\bibitem{Ban93}
S. Banerjee $et~al.$, Phys. Lett. {\bf B305}, 182 (1993).

\bibitem{Bel02}
 A. Belogianni $et~al.$,Phys. Lett. {\bf B548}, 122 (2002).

\bibitem{Bel02a}
 A. Belogianni $et~al.$,Phys. Lett. {\bf B548}, 129 (2002).

\bibitem{DEL06}
J. Abdallah $et~al.$ (DELPHI Collaboration), Eur.  Phys. J. {\bf C47}, 273
(2006).  

\bibitem{DEL08}
 J. Abdallah $et~al.$ (DELPHI Collaboration), Eur.  Phys. J. {\bf C57}, 499 
(2008).

\bibitem{DEL10}
 J. Abdallah $et~al.$ (DELPHI Collaboration),
Eur. Phys. J. {\bf C67}, 343 (2010).

\bibitem{Per09} V. Perepelitsa, for the DELPHI Collaboration,
Nonlinear Phenomena
  in Complex Systems, {\bf 12}, 343 (2009).

\bibitem{Won10}
C. Y. Wong, Phy.  Rev. {\bf C81}, 064903 (2010).

\bibitem{Van89}
L. Van Hove, Ann. Phys. (N.Y.)  {\bf 192}, 66 (1989);
V. Balek, N. Pisutova, and J. Pisut,
Acta.  Phys. Pol. {\bf B21}, 149 (1990);
W. Czyz and W. Florkowski, Z. Phys. {\bf C61}, 171 (1994);
P. Lichard, Phys. Rev. {\bf D50}, 6824 (1994);
 E. Kokoulina $et~al.$, Braz. J.
Phys., {\bf 37}, 785 (2007); S.M. Darbinian $et~al.$,
 Sov. J. Nucl. Phys. {\bf 54}, 364 (1991);
 O. Nachtmann, hep-ph/9411345;
Yu.A. Simonov, A.I. Veselov, Phys. Lett.
{\bf B671}, 55 (2009);
Y. Hatta and T. Ueda, Nucl.Phys. {\bf B837}, 22 (2010);
L. Labun and J. Rafelski, arXiv:1010.1970 (2010).

\bibitem{Col75}
S. Coleman, R. Jackiw, and L. Susskind, Ann. Phys. {\bf 93}, 267 (1975);
S. Coleman, Ann. Phys. {\bf 101}, 239 (1976).

\bibitem{Wit84}
E. Witten, Commun. Math. Phy. {\bf 92}, 455 (1984).

\bibitem{Sch62}
J. Schwinger, Phys. Rev. {\bf 128}, 2425 (1962);
J. Schwinger, in $Theoretical$ $Physics$, Trieste Lectures, 1962
  (I.A.E.A., Vienna, 1963), p. 89.


\bibitem{Bar97} R. Barate $et~al.$ (ALEPH Collaboration),
Zeit. Phys. {\bf C74}, 451 (1997).


\bibitem{Cas74} 
A. Casher, J. Kogut, and  L.  Susskind,  Phys.  Rev.  {\bf D10},  732
(1974).

\bibitem{Won09} 
C. Y. Wong, Phys. Rev. {\bf C80}, 034908 (2009). 

\bibitem{Won09a} 
C. Y. Wong, Phys. Rev. {\bf C80}, 054917 (2009). 


\end{thebibliography}




\end{document}